\documentclass[a4paper]{article}

\usepackage{INTERSPEECH2022}
\usepackage{amsmath,graphicx}
\usepackage{subcaption}
\usepackage{booktabs}
\usepackage{multirow}
\usepackage{makecell}
\usepackage{tabularx}
\usepackage{array}
\usepackage{bm}
\usepackage{hyperref}

\def\our{GlowVC}
\def\ourE{\our{}-explicit}
\def\ourEcell{\makecell{\our{}-\\explicit}}
\def\ourC{\our{}-conditional}
\def\ourCcell{\makecell{\our{}-\\cond}}
\newcolumntype{L}{>{\raggedleft\arraybackslash}X}
\newcolumntype{R}{>{\raggedright\arraybackslash}X}
\newcolumntype{C}{>{\centering\arraybackslash}X}

\title{\our{}: Language-independent text-free voice conversion \\ and the mel-spectrogram space disentangling model}
\title{\our{}: Mel-spectrogram space disentangling model \\ for language-independent text-free voice conversion}
\name{Magdalena Proszewska$^1$$^*$\thanks{$^*$This research was completed during an internship at Alexa AI.}, Grzegorz Beringer$^2$, Daniel Sáez-Trigueros$^2$, Thomas Merritt$^2$, Abdelhamid Ezzerg$^2$, Roberto Barra-Chicote$^2$}
\address{
  $^1$Jagiellonian University, Poland\\
  $^2$Alexa AI}
\email{magdalena.proszewska@student.uj.edu.pl, \{beringg,dsaez,thommer,ezzerg,rchicote\}@amazon.com}

\begin{document}
\ninept

\maketitle

\begin{abstract}

In this paper, we propose \our{}: a multilingual multi-speaker flow-based model for language-independent text-free voice conversion.
We build on Glow-TTS, which provides an architecture that enables use of linguistic features during training without the necessity of using them for VC inference.
We consider two versions of our model: \ourC{} and \ourE{}.
\ourC{} models the distribution of mel-spectrograms with speaker-conditioned flow and disentangles the mel-spectrogram space into content- and pitch-relevant dimensions, while \ourE{} models the explicit distribution with unconditioned flow and disentangles said space into content-, pitch- and speaker-relevant dimensions.
We evaluate our models in terms of intelligibility, speaker similarity and naturalness for intra- and cross-lingual conversion in seen and unseen languages.
\our{} models greatly outperform AutoVC baseline in terms of intelligibility, while achieving just as high speaker similarity in intra-lingual VC, and slightly worse in the cross-lingual setting.
Moreover, we demonstrate that \ourE{} surpasses both \ourC{} and AutoVC in terms of naturalness.

\end{abstract}

\vspace{-2mm}
\section{Introduction}
\label{sec:intro}

The goal of voice conversion (VC) is to change the speaker identity of the input speech to another speaker, while maintaining the linguistic content, intelligibility and naturalness.
Non-parallel text-free VC assumes we have no paired samples to learn how to change the speaker identity in a supervised fashion, and no input text provided during inference.

Recently, due to the advancements in the deep learning field, several non-parallel text-free VC models, which achieve high naturalness and target speaker similarity, have been proposed \cite{vcc2020}.
ASR-based approaches first use a pre-trained ASR model to extract the speaker-invariant linguistic features, and then build a VC model that takes these features and speaker embeddings as input.
The extracted features can be either the recognized text (i.e. ASR+TTS pipeline) \cite{ASR-TTS,Cotatron} or frame-level phonetic information, e.g. Phoneme Posteriorgrams (PPGs) \cite{AverageModeling,BilingualPPGs}.
In GAN-based approaches, the generator network and the discriminator network are trained jointly and generator is encouraged by the adversarial loss from discriminator to produce speech indistinguishable from real recordings of the target speaker, e.g. StarGAN-VC \cite{StarGAN-VC,Star-GAN-VC2}.
Autoencoder-based approaches aim to disentangle speech into speaker and linguistic information during reconstruction training and perform VC by changing the speaker information during inference \cite{qian2019autovc,chou2019one,VQVAE}.
Finally, it is possible to leverage TTS for improving the VC model, e.g. pre-training the VC decoder as TTS \cite{VTN} or using a multi-task approach and training both TTS and VC at the same time \cite{Nautilus}.

While above methods usually work well for intra-lingual VC (i.e. source and target speak the same language), they can struggle in achieving similar level of naturalness and speaker similarity in cross-lingual VC \cite{vcc2020}.
Moreover, some of these methods do not scale well as we increase the number of languages that we want to support in a single VC model.
ASR-based approaches would require either training separate ASR models for each language \cite{BilingualPPGs,CrossLingualCycleConsistencyPPGs}, or training a multilingual phoneme recognizer \cite{UniversalLinguisticRepresentations}.
Autoencoders need to enforce a strong information bottleneck in order to reduce speaker leakage from the encoder \cite{qian2019autovc}, often suffering from low intelligibility at the cost of acceptable target speaker similarity.
Finally, one might want to support unseen languages at inference time, without the need to re-train the VC system on new data.
In this case, language-independence and good generalization is required.

We propose \our{}: a non-parallel language-independent text-free any-to-any VC system built on top of Glow-TTS \cite{kim2020glowtts}.
Due to its design and invertibility of the flow decoder, we can train the model with text (TTS training), but perform text-free VC at inference time (VC inference) \cite{kim2020glowtts,casanova2021scglowtts}.
Two variants of \our{} are demonstrated:
\textit{\ourC{}}, that conditions the flow decoder on speaker identity and performs VC by changing said identity at decoding stage (similarly to original Glow-TTS \cite{kim2020glowtts});
\textit{\ourE{}}, that explicitly disentangles speech into content, pitch and speaker information and performs VC by manipulating the speaker-relevant latents, without the need for any conditioning in the decoder.

We perform experiments with intra- and cross-lingual VC in seen and unseen languages, and evaluate our models with objective ntelligibility metric (WER), as well as subjective metrics: MUSHRA speaker similarity and naturalness \cite{itu20031534}.
We show that both \our{} models significantly outperform the AutoVC baseline \cite{qian2019autovc} in intelligibility and perform well even if the language of the source sample was not seen during training.
Moreover, explicit disentanglement of \ourE{} is shown to significantly improve naturalness, outperforming both \ourC{} and AutoVC.
However, speaker similarity for cross-lingual VC is noticeably lower than for intra-lingual VC and overall worse than for AutoVC.
We leave improvements on speaker similarity for future work.

\vspace{-3mm}
\section{GlowVC} \label{sec:models}

\our{} models are based on the Glow-TTS \cite{kim2020glowtts}. Unlike in Glow-TTS, instead of jointly training phoneme alignments with monotonic alignment search (MAS), we use pre-trained explicit phoneme durations (similarly to attention-free TTS \cite{shah21_ssw}), simplifying and accelerating the training process.
In our internal tests, we found models with MAS to be difficult to train, often leading to subpar alignments that impacted the naturalness of the system.

Furthermore, content encoder (originally text encoder) architecture is simplified and extended with language information, and normalized $F_0$ is added to model pitch.
The use of explicit durations and $F_0$ allow our models to achieve more natural prosody than shown in \cite{kim2020glowtts,casanova2021scglowtts}.

\vspace{-3mm}
\subsection{\ourC{}} \label{subsec:glow-tts}
\ourC{} is a multilingual multi-speaker version of Glow-TTS that models the conditional distribution of mel-spectro-grams $x$ as $P_X(x|c,p,s)$, where $c$ denotes content information (text, language and durations), $p$ denotes pitch information (normalized $F_0$) and $s$ -- speaker information (speaker embeddings).
It transforms the conditional prior distribution $P_Z(z|c,p)$ through an invertible, flow-based decoder $f^{(s)}:z\to x$ that is conditioned on $s$ and disentangles the mel-spectrogram signal into two speech factors: content and pitch.
The architecture is shown in Figure \ref{fig:glow-tts}.

Here, $Z$ consists of two independent random variables $Z=(Z^{(c)}, Z^{(p)})$ that correspond to content and pitch, and have isotropic multivariate Gaussian distributions $\mathcal{N}(\mu^{(c)}, \sigma^{(c)})$ and $\mathcal{N}(\mu^{(p)}, \sigma^{(p)})$, respectively.
Let $T$ denote length of mel-spectrogram.
The parameters $\mu^{(c)}=\mu^{(c)}_{1:T}$, $\sigma^{(c)}=\sigma^{(c)}_{1:T}$ are matrices obtained from content encoder, while $\mu^{(p)}=\mu^{(p)}_{1:T}=p$, $\sigma^{(p)}=1$ are vectors obtained directly from $p$.
By the change of variables
\begin{equation}
\log P_X(x|c,p,s) = \log P_Z(z|c,p) + \log \Bigg\vert det \frac{\partial (f^{(s)})^{-1}(x)}{\partial x}\Bigg\vert,
\end{equation}
where
\begin{align}
\label{eq:logpz_conditional}
\begin{split}
 \log P_Z(z|c,p)	= \log P_{Z^{(c)}}(z^{(c)}|c) +\log P_{Z^{(p)}}(z^{(p)}|p)
\end{split}
\end{align}
and
\vspace{-2mm}
\begin{equation}
\label{eq:logpz_feature}
	\log P_{Z^{(d)}}(z^{(d)}|d) = \sum_{j=1}^T \log \mathcal{N}(z^{(d)}_j ;\mu^{(d)}_j, \sigma^{(d)}_j)
\end{equation}
for any feature $d$.
The objective is given by
\begin{equation}
\label{eq:loss}
	\max L(\theta) = \max_\theta \log P_X(x|c,p,s;\theta),
\end{equation}
where $\theta$ parametrises the network.

During TTS inference, statistics of the prior distribution are determined by content encoder and normalized $F_0$.
Then, a latent variable is sampled from the prior distribution and transformed into a mel-spectrogram using the decoder.

Voice conversion from the source speaker $s_0$ to the target speaker $s_1$ is defined as $f^{(s_1)}\circ (f^{(s_0)})^{-1}$.
Only the decoder is used and the content and pitch information is not required.
The approach of using the condition to first remove the source speaker characteristics when going from $x$ to $z$, and then imprint the target speaker characteristics during inverse transformation, has been also explored in other flow-based TTS and VC models \cite{kim2020glowtts,casanova2021scglowtts, Blow}.

\subsection{\ourE{}}
We propose an alternative model, \ourE{}, which learns to disentangle the mel-spectrogram signal into three speech factors: content, pitch and speaker identity.
As in \ourC{}, we model the conditional distribution $P_X(x|c,p,s)$.
However, here, Z consists of three independent random variables $Z=(Z^{(c)}, Z^{(p)},Z^{(s)})$ and the conditional prior distribution $P_Z(z|c,p,s)$ is transformed through an invertible, flow-based decoder $f: z\to x$.
The architecture of \ourE{} is shown in Figure \ref{fig:glow-tts-prior}.

The prior distributions $P_{Z^{(c)}}$, $P_{Z^{(p)}}$ are defined as in Section \ref{subsec:glow-tts}.
The prior distribution $P_{Z^{(s)}}$ is the isotropic multivariate Gaussian distribution $\mathcal{N}(\mu^{(s)}, \sigma^{(s)})$, in which the parameters $\mu^{(s)}$ and $\sigma^{(s)}$ are broadcasted vectors obtained from the speaker encoder.
Equation \eqref{eq:logpz_conditional} is now of the form:

\vspace{-4mm}
\begin{align}
\label{eq:logpz_explicit}
\begin{split}
\log P_Z(z|c,p,s) = \log P_{Z^{(c)}}(z^{(c)}|c) \\ + \log P_{Z^{(p)}}(z^{(p)}|p) \\
	+ \log P_{Z^{(s)}}(z^{(s)}|s)
\end{split}
\end{align}
As in \ourC{}, during TTS inference, a latent variable is sampled from the prior distribution and transformed into a mel-spectrogram.

In order to perform voice conversion, we encode utterance of the source speaker as $z=f^{-1}(x)$, keep latents $z^{(c)}, z^{(p)}$ unchanged and switch latent $z^{(s)}$ to broadcasted $\mu^{(s)}$ obtained by encoding the target speaker with the speaker encoder. Then, we decode modified $z$ and obtain output utterance. Unlike \ourC{}, \ourE{} does not require source speaker information for VC inference.

\vspace{-2mm}
\subsection{Model architecture}

\begin{figure*}[htb]
\begin{subfigure}{.33\textwidth}
  \centering
  \includegraphics[height=135px]{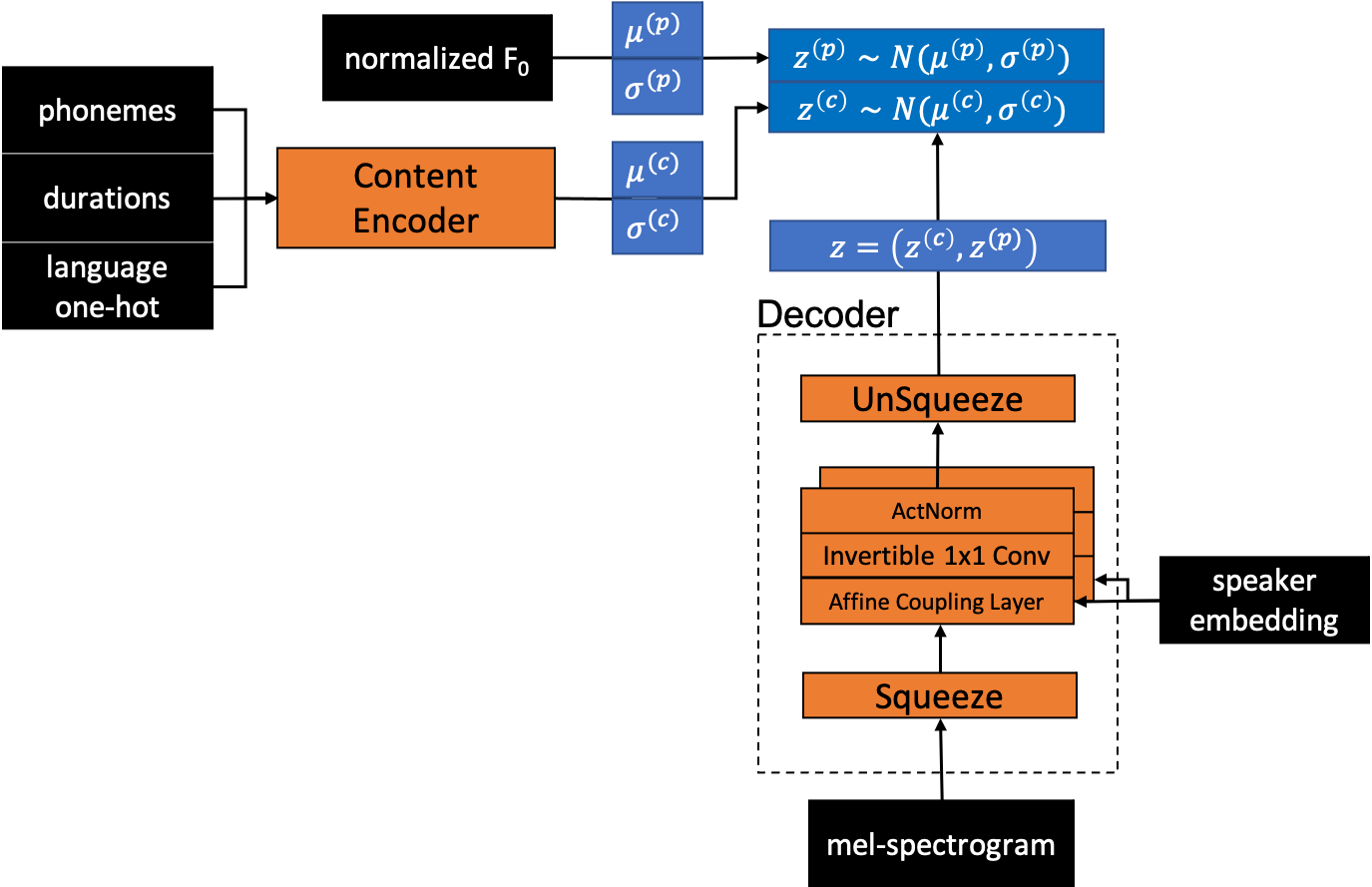}
  \caption{\ourC{}}
  \label{fig:glow-tts}
\end{subfigure}%
\hspace{10mm}
\begin{subfigure}{.33\textwidth}
   \includegraphics[height=135px]{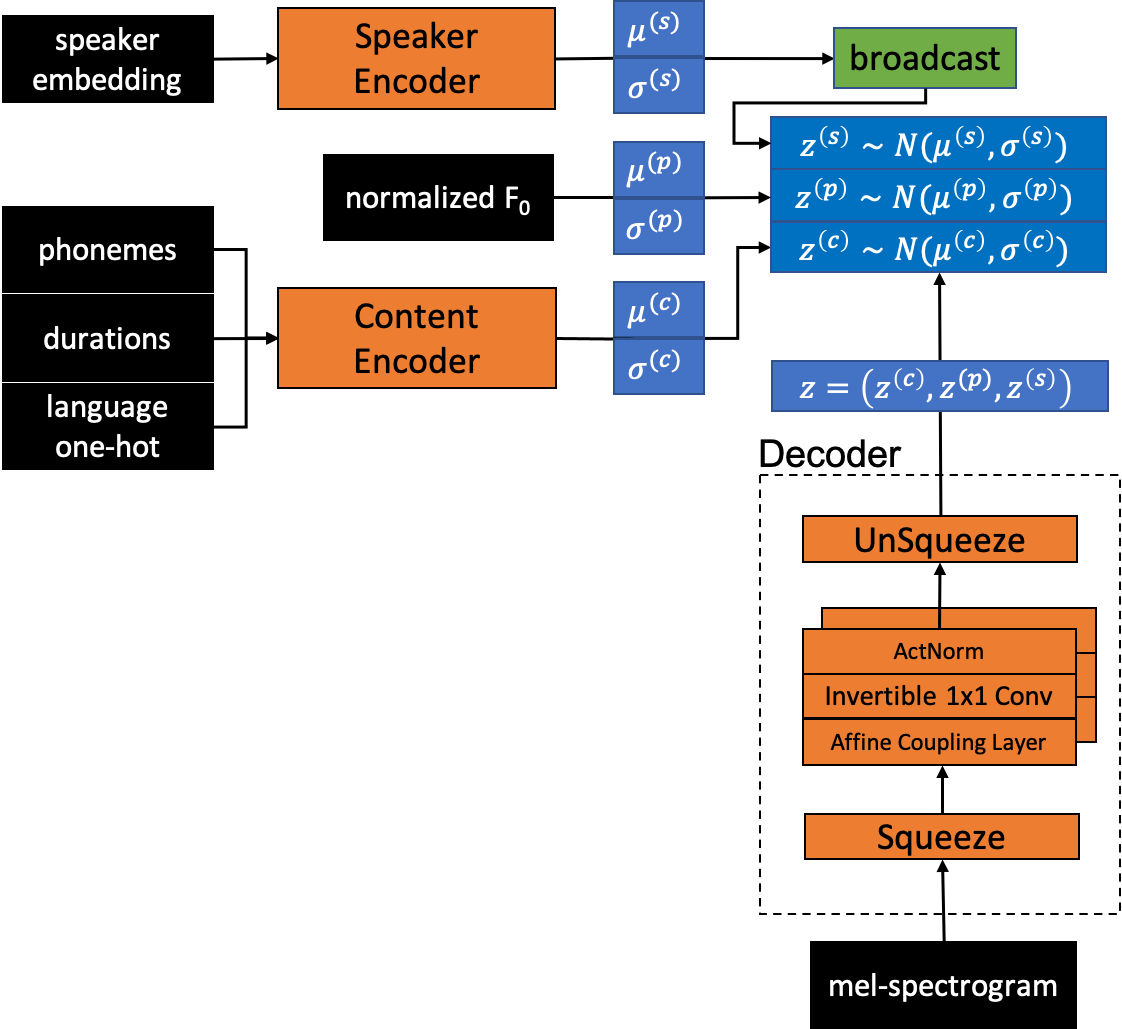}
   \caption{\ourE{}}
  \label{fig:glow-tts-prior}
\end{subfigure}
\begin{subfigure}{.33\textwidth}
	\hspace{7mm}
   \includegraphics[height=135px]{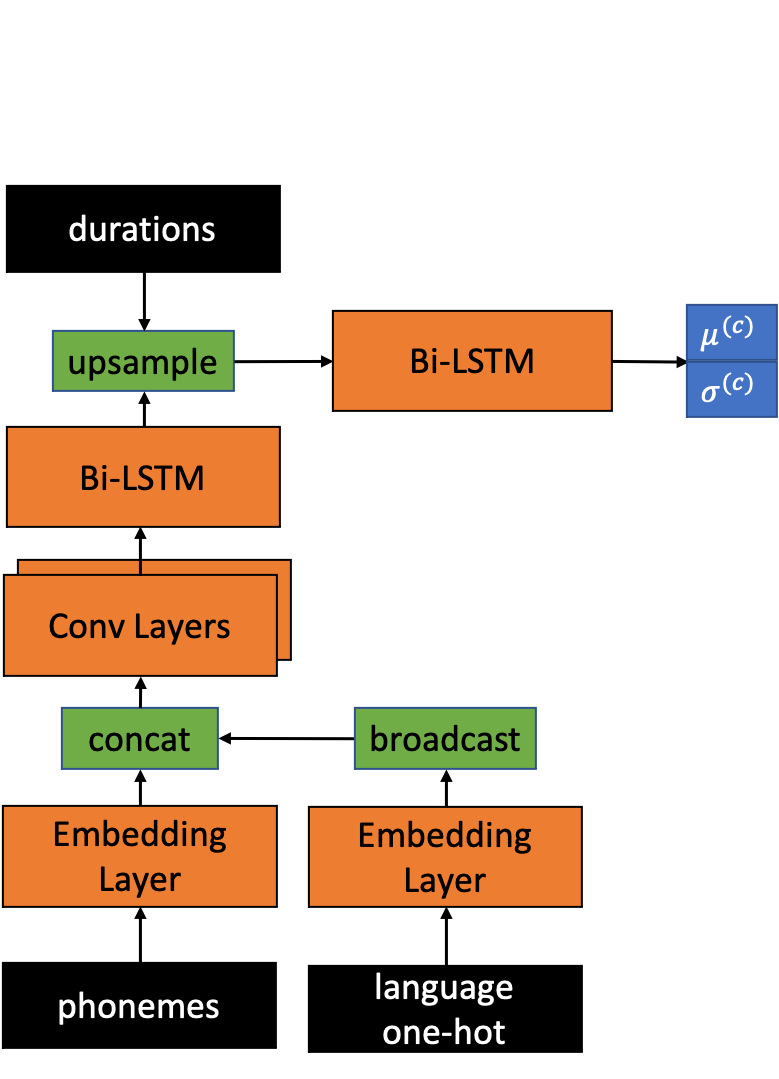}
   \caption{Content encoder}
   \label{fig:content-encoder}
\end{subfigure}
\caption{Architecture of our flow-based models. In VC inference, only decoder and speaker encoder are used.}
\label{fig:models}
\end{figure*}

Both \our{} models take mel-spectrograms, speaker embeddings, normalized $F_0$, phonemes, language one-hots and durations as inputs.
Except mel-spectrograms and speaker embeddings, all the other inputs are only used to determine the parameters of the flow prior during training and, therefore, are not required during VC inference.
For calculating mean of the prior, we use content encoder and additional speaker encoder in \ourE{}.
As in \cite{kim2020glowtts}, standard deviation for all priors is set to $1$.

\vspace{-4mm}
\paragraph*{Content encoder}
Instead of the encoder proposed in \cite{kim2020glowtts} (based on TransformerTTS \cite{transformerTTS}), we use a simpler architecture.
Phonemes and language label are first embedded, then concatenated and passed through the convolutional layers and the first BiLSTM layer.
The output features are upsampled according to given phoneme durations and finally passed through the second BiLSTM layer with the purpose of smoothing the prior.
For our experiments, we used $4$ conv layers with $512$ channels, kernel size of $5$, batch normalization, ReLU activation and dropout of $0.2$.
Hidden size of BiLSTM layers is set to the size of content-based prior.
Content encoder is shown in Figure \ref{fig:content-encoder}.

\vspace{-4mm}
\paragraph*{Speaker encoder}
Speaker encoder of \ourE{} is a linear layer that projects dimensionality of speaker embeddings to the size of chosen speaker-based prior.

\vspace{-4mm}
\paragraph*{Flow-based decoder}
Decoders in both \our{} models are based on Glow \cite{kingma2018glow} and conditional Glow (as in \cite{kim2020glowtts}), respectively.
Their architecture and parameters are defined as in the original Glow-TTS implementation \cite{kim2020glowtts} except that the hidden channels number in \ourE{} is increased and set to $384$.
\ourE{} learns a more difficult task than \ourC{} ie. extracting speaker embeddings, not processing them, and we found that the higher hidden channel size is important for a stable performance.

\vspace{-3mm}
\section{Experiments} \label{sec:experiments}

\vspace{-2mm}
\subsection{Dataset}
All models were trained on a multi-lingual clean-speech dataset recorded in studio-quality conditions with professional voice actors.
Overall, 380 hours of recordings (280k utterances) in 5 languages (de-DE, en-US, es-ES, fr-FR and it-IT) were used.
For evaluation we used unseen utterances from all 25 seen speakers, as well as utterances from additional 6 speakers of 3 unseen languages: en-GB, pt-BR and ru-RU (2 speakers per language).
Number of hours and speakers per language available in the training set are shown in Table \ref{tab:data}.

\addtolength{\tabcolsep}{-2.5pt}
\begin{table}[htb]
	\centering
	\caption{Details of the training set.}
	\label{tab:data}
	\footnotesize
	\begin{tabularx}{0.75\linewidth}{cccccccccc}
	\toprule
	Language & de-DE & en-US & es-ES & fr-FR & it-IT \\
	\midrule
	\makecell{\# hours} & 89 & 142 & 52 & 44 & 51 \\
	\midrule
	\makecell{\# speakers} & 4 & 10 & 4 & 4 & 3  \\
	\bottomrule
	\end{tabularx}
\end{table}
\addtolength{\tabcolsep}{2.5pt}

Audio data was sampled at 16kHz and processed into $80$ dimensional mel-spectrograms with 50ms frames and 12.5ms frame-shift.
For modeling pitch, we used interpolated log-normalized $F_0$, which was standardized at the utterance level to prevent speaker leakage.
Phonemes were extracted from text using a propietary linguistic front-end.
Durations were extracted with an external Gaussian Mixture Model (GMM) aligner trained with Kaldi Speech Recognition Toolkit \cite{Povey_ASRU2011}, similarly to \cite{shah21_ssw}.
Speaker embeddings of size 192 were extracted using the flow-based speaker encoder from \cite{vallesperez2021improving}.
For mel-spectrogram to wave synthesis, Parallel WaveNet universal vocoder was used \cite{jiao2021universal}.

\subsection{Models and training parameters}

\paragraph*{\our{}}
Input and output of the flow decoder must be of the same shape, hence they both have 80 dimensions (size of mel-spectrograms).
In \ourC{}, $z$ is split into (79, 1) dimensions for \textit{(content, pitch)}.
In \ourE{}, $z$ is split into (40, 39, 1) dimensions for \textit{(content, speaker, pitch)}.
These values were chosen arbitrarily - we leave the impact analysis of content/speaker/pitch dimensionalities for future work.
For training of the models, we use batch size of 32, Adamax optimizer with learning rate of 0.0001 and a warm-up schedule of 5 epochs.
Models were trained for 350k steps on 2 NVIDIA V100 GPUs.

\vspace{-4mm}
\paragraph*{AutoVC baseline}
Since we are interested in evaluating cross-lingual VC for both seen and unseen languages, we are restricted to a very narrow family of language-independent VC models.
Autoencoder-based models fit the requirements, as they do not need text or language-specific information at inference time, unlike ASR- or PPG-based approaches.
Specifically, we choose AutoVC \cite{qian2019autovc} for our baseline, due to its popularity in VC community.
Compared to the original version, we concatenate speaker embeddings to input spectrogram in order to reduce speaker leakage as in CopyCat \cite{Karlapati_2020}.
We also use a bottleneck of 8 and downsampling frequency of 16, which we found to be crucial in order to preserve content and get good target speaker similarity.
AutoVC was trained with batch size of 32 and Adam optimizer with learning rate of 0.001 for 200k steps on 1 NVIDIA V100 GPU.

\subsection{Experimental setup}
We carried out 4 experiments to measure VC performance of \our{} models against AutoVC:
\begin{enumerate}
	\item Intra-lingual VC between seen speakers (seen $\to$ seen);
	\item Cross-lingual VC between seen speakers (seen $\to$ seen);
	\item Cross-lingual VC from seen speakers to unseen speakers of unseen languages (seen $\to$ unseen);
	\item Cross-lingual VC from unseen speakers of unseen languages to seen speakers (unseen $\to$ seen).
\end{enumerate}

\vspace{-2mm}
\subsection{Evaluation metrics}
The VC models were evaluated in terms of speech intelligibility and naturalness, and speaker similarity.

\vspace{-3mm}
\paragraph*{Word Error Rate (WER)}
WER is an objective metric used to measure intelligibility of speech.
The VC samples are transcribed with AWS Transcribe\footnote{https://aws.amazon.com/transcribe/}, and compared to the original transcriptions.

\vspace{-3mm}
\paragraph*{MUSHRA speaker similarity}
We measure subjective speaker similarity with a MUSHRA perceptual speech test.
This test consists of 800 test cases split into 4 experiments: intra-lingual (250), cross-lingual between seen speakers (250), cross-lingual from unseen speakers and languages (150), cross-lingual to unseen speakers and languages (150).
Source speaker, their utterance and target speaker were selected randomly to ensure an equal representation of gender and language transformations.
Each test case was presented to 20 listeners, native in the source speaker language.
In each test, listeners were provided with a reference audio from target speaker and 5 audios to be assessed: one from each of the 3 models, the unmodified source audio and a random audio from the target speaker.
They were asked to rate samples based on their voice identity similarity to the target speaker on a scale from 0 to 100.

\vspace{-3mm}
\paragraph*{MUSHRA Naturalness}
We measure subjective naturalness of converted samples with a MUSHRA test, similar to the one described above.
For this test, listeners were provided only with 4 audios to be assessed: one from each of the 3 models and the reference audio - a random audio in the source language, from speaker of the same gender as the target speaker (if possible).
The listeners were asked to rate each model on a scale from 0 to 100 in terms of naturalness.
\\
\\
We perform paired t-tests with Holm-Bonferroni correction to detect statistically significant differences between systems in MUSHRA tests. All reported significant
differences are for $p < 0.05$.

\addtolength{\tabcolsep}{-1.5pt}
\begin{table}[htb]
	\centering
	\caption{WER (\%) results for intra- and cross-lingual VC for seen speakers and cross-lingual VC from seen to unseen speakers (experiments 1-3). GT - ground truth, S - seen, U - unseen.}
	\footnotesize
	\vspace{-3mm}
	\begin{tabularx}{0.95\linewidth}{cccccc}
		\toprule
		Experiment & GT  & AutoVC &  \ourCcell{} & \ourEcell{} \\
		\midrule
		\makecell{intra-lingual (S $\to$ S)} & 11.56 & 22.44 & \textbf{12.44} & 12.52 \\
		\midrule
		\makecell{cross-lingual (S $\to$ S)} & - & 40.10 & 14.63 & \textbf{14.11} \\
		\midrule
		\makecell{cross-lingual (S $\to$ U)} & - & 36.93 & 14.33 & \textbf{13.67} \\
		\bottomrule
	\end{tabularx}
	\label{tab:cross-ligual}
\end{table}
\addtolength{\tabcolsep}{1.5pt}

\vspace{-5mm}

\addtolength{\tabcolsep}{-0.5pt}
\begin{table}[htb]
	\centering
	\caption{WER (\%) results for cross-lingual VC from unseen to seen speakers (experiment 4).}
	\footnotesize
	\vspace{-3mm}
	\begin{tabularx}{0.84\linewidth}{cccccc}
		\toprule
		Language & GT  & AutoVC &  \ourCcell{} & \ourEcell{} \\
		\midrule
		en-GB & 7.77 & 41.75 & \textbf{8.90} & 9.29 \\
		\midrule
		pt-BR & 16.74 & 45.65 & \textbf{22.45} & 23.04 \\
		\midrule
		ru-RU & 29.74 & 55.82 & 37.50 & \textbf{37.36}\\
		\bottomrule
	\end{tabularx}
	\label{tab:zero-shot}
\end{table}
\addtolength{\tabcolsep}{0.5pt}

\begin{figure}[htb!]
\begin{subfigure}{0.48\textwidth}
  \centering
  \includegraphics[width=1\linewidth]{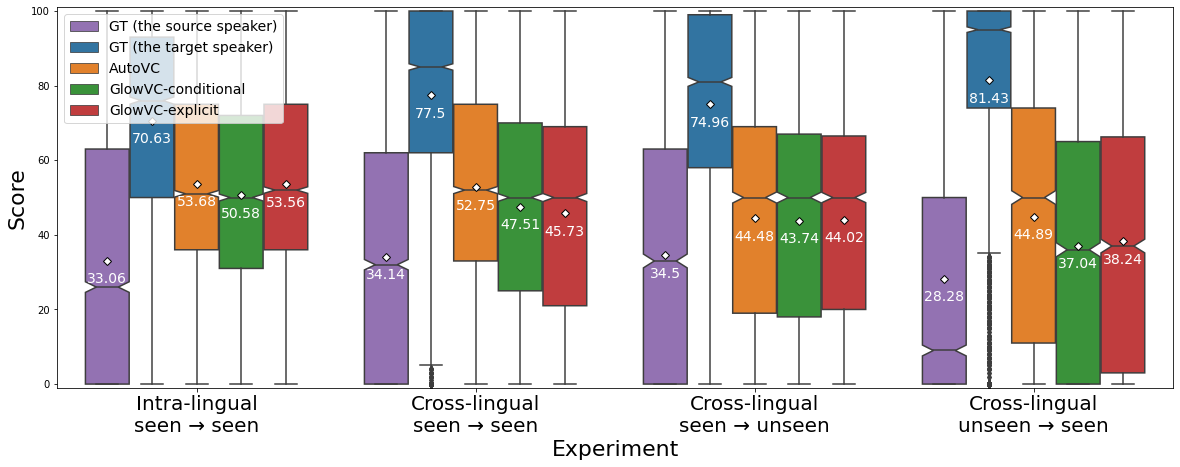}
  \caption{Speaker similarity}
  \label{fig:speaker-similarity}
\end{subfigure}
\begin{subfigure}{0.48\textwidth}
  \centering
  \includegraphics[width=1\linewidth]{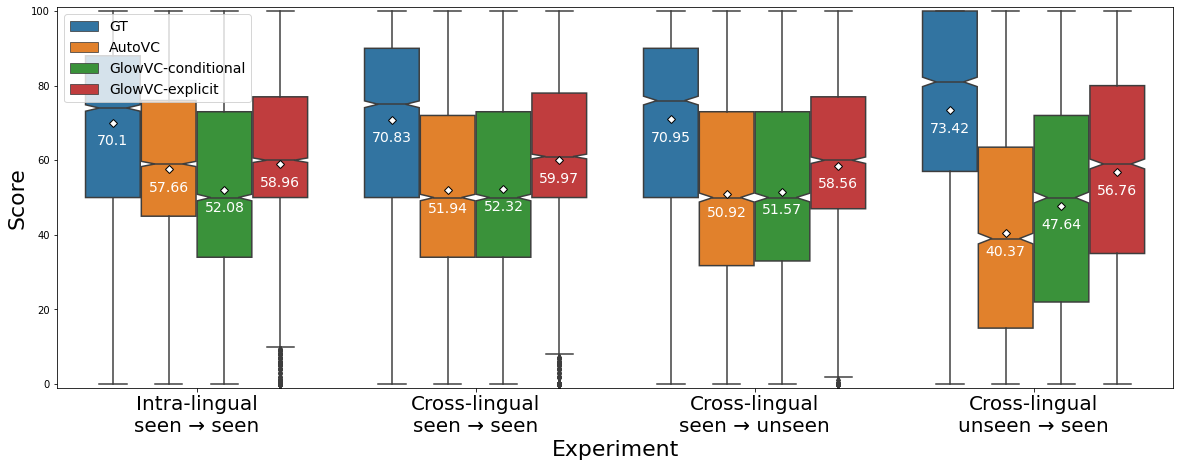}
  \caption{Naturalness}
  \label{fig:naturalness}
\end{subfigure}
\caption{Human scores from MUSHRA tests.}
\vspace{-4mm}
\label{fig:metrics}
\end{figure}

\vspace{-4mm}
\subsection{Results and discussion}
Objective WER results can be seen in Tables \ref{tab:cross-ligual} and \ref{tab:zero-shot}.
Subjective MUSHRA results can be seen in Figure \ref{fig:metrics}.

The biggest advantage of \our{} is the ability to produce highly intelligible speech without using content information during inference, even in languages unseen during training.
This can be seen in \our{}'s WER results being very close to ground-truth WER, while AutoVC substantially increases WER, especially when doing cross-lingual VC (Table \ref{tab:cross-ligual}, \ref{tab:zero-shot}).

MUSHRA speaker similarity evaluations (Figure \ref{fig:speaker-similarity}) find that AutoVC and \ourE{} are rated as significantly more similar to the target speaker than \ourC{} for intra-lingual VC, and that AutoVC outperforms both \our{} models at cross-lingual VC (excluding "seen $\to$ unseen" scenario where there were no significant differences found between the models).
We believe that AutoVC is better at cross-lingual speaker similarity since it often copies the accent of the target speaker onto the source sample, which boosts similarity scores at the cost of poor intelligibility.
Let us consider an example of converting an English speaker (source) to a German one (target).
We found that samples converted with AutoVC have a very strong German accent, which severely impacts intelligibility (Table \ref{tab:cross-ligual}) and naturalness (\ref{fig:naturalness}) but does improve speaker similarity scores, since accent plays a part in perceived speaker identity.
On the other hand, \our{} models learn to remove language information from speaker embeddings and do not change the accent of the source utterance when performing speaker conversion.

In terms of naturalness, \ourE{} significantly outperforms the other two models in all MUSHRA tests (Figure \ref{fig:naturalness}), which we attribute to high intelligibility and the ability to preserve the accent of the source utterance during cross-lingual VC.
Interestingly, \ourC{} is rated significantly worse compared to \ourE{}.
Upon listening to evaluation samples, we attribute that difference to the fact that samples from \ourC{} have worse signal quality compared to \ourE{}.
We hypothesize that transitioning between speakers via latent $z$ is smoother than with conditional flow, hence converted samples sound smoother and more human-like with \ourE{}.

\vspace{-2mm}
\section{Conclusions and future work}
We presented \our{}: a language-independent voice conversion model that demonstrates great potential at maintaining linguistic content, intelligibility and naturalness, even for unseen languages.
We considered two versions of the model: \ourC{} and \ourE{}.
We tested their abilities in intra- and cross-lingual VC scenario for seen and unseen languages, and showcased the high level of intelligibility achieved by the proposed models.
Finally, we showed that \ourE{} performs better in terms of naturalness than \ourC{}, does not require source speaker information for VC inference and allows for disentanglement of mel-spectrogram into content-, pitch- and speaker-relevant dimensions.
In future work, we intend to investigate the use of \ourE{} for modifying pitch and speech rate, extracting speaker embeddings and performing smooth speaker interpolations between speakers.

\clearpage
\bibliographystyle{IEEEtran}
\bibliography{refs}

\end{document}